\documentclass[preprintnumbers,amsmath,amssymbm,prd]{revtex4}
\usepackage{epsfig}
\usepackage{graphicx}

\begin{document}
\title{A lower bound on the Bekenstein-Hawking temperature of black holes}
\author{Shahar Hod}
\address{The Ruppin Academic Center, Emeq Hefer 40250, Israel}
\address{ }
\address{The Hadassah Institute, Jerusalem 91010, Israel}
\date{\today}

\begin{abstract}
\ \ \ We present evidence for the existence of a quantum lower bound
on the Bekenstein-Hawking temperature of black holes. The suggested
bound is supported by a gedanken experiment in which a charged
particle is dropped into a Kerr black hole. It is proved that the
temperature of the final Kerr-Newman black-hole configuration is
bounded from below by the relation $T_{\text{BH}}\times
r_{\text{H}}>(\hbar/r_{\text{H}})^2$, where $r_{\text{H}}$ is the
horizon radius of the black hole.
\end{abstract}
\bigskip
\maketitle

\section{Introduction}

It is well known \cite{Bek1,Lan} that, for mundane physical systems
of spatial size $R$, the thermodynamic (continuum) description
breaks down in the low-temperature regime $T\sim \hbar/R$
\cite{Noteunit}. In particular, these low temperature systems are
characterized by thermal fluctuations whose wavelengths
$\lambda_{\text{thermal}}\sim \hbar/T$ are of order $R$, the spatial
size of the system, in which case the underlying quantum (discrete)
nature of the system can no longer be ignored. Hence, for mundane
physical systems of spatial size $R$, the physical notion of
temperature is restricted to the high-temperature thermodynamic
regime \cite{Bek1,Lan}
\begin{equation}\label{Eq1}
T\times R\gg \hbar\  .
\end{equation}

Interestingly, black holes are known to have a well-defined notion
of temperature in the complementary regime of low temperatures. In
particular, the Bekenstein-Hawking temperature of generic
Kerr-Newman black holes is given by \cite{Bek2,Haw}
\begin{equation}\label{Eq2}
T_{\text{BH}}={{\hbar(r_+-r_-)}\over{4\pi (r^2_++a^2)}}\  ,
\end{equation}
where
\begin{equation}\label{Eq3}
r_{\pm}=M+(M^2-a^2-Q^2)^{1/2}
\end{equation}
are the radii of the black-hole (outer and inner) horizons
\cite{Notema}. The relation (\ref{Eq2}) implies that near-extremal
black holes in the regime $(r_+-r_-)/r_+\ll1$ are characterized by
the strong inequality \cite{Notethr}
\begin{equation}\label{Eq4}
T_{\text{BH}}\times r_+\ll \hbar\  .
\end{equation}

It is quite remarkable that black holes have a well defined notion
of temperature in the regime (\ref{Eq4}) of low temperatures, where
mundane physical systems are governed by finite-size (quantum)
effects and no longer have a self-consistent thermodynamic
description.

One naturally wonders whether black holes can have a physically
well-defined notion of temperature all the way down to the extremal
({\it zero}-temperature) limit $T_{\text{BH}}\times r_+/\hbar\to 0$?
In order to address this intriguing question, we shall analyze in
this paper a gedanken experiment which is designed to bring a
Kerr-Newman black hole as close as possible to its extremal limit.
We shall show below that the results of this gedanken experiment
provide compelling evidence that the Bekenstein-Hawking temperature
of the black holes is bounded from below by the quantum inequality
$T_{\text{BH}}\times r_+\gg (\hbar/r_+)^2$.

\section{The gedanken experiment}

We consider a spherical body of proper radius $R$, rest mass $\mu$,
and electric charge $q$ which is slowly lowered towards a Kerr black
hole of mass $M$ and angular momentum $J=Ma$ along the symmetry axis
of the black hole \cite{Notegen}. The black-hole spacetime is
described by the line element \cite{Chan,Kerr,NoteBL}
\begin{eqnarray}\label{Eq5}
ds^2=-{{\Delta}\over{\rho^2}}(dt-a\sin^2\theta
d\phi)^2+{{\rho^2}\over{\Delta}}dr^2+\rho^2
d\theta^2+{{\sin^2\theta}\over{\rho^2}}\big[a
dt-(r^2+a^2)d\phi\big]^2\  ,
\end{eqnarray}
where $\Delta\equiv r^2-2Mr+a^2$ and $\rho^2\equiv
r^2+a^2\cos^2\theta$.

The test-particle approximation implies that the parameters of the
body are characterized by the strong inequalities
\begin{equation}\label{Eq6}
\mu\ll R\ll r_+\  .
\end{equation}
These relations imply that the particle which is lowered into the
black hole has negligible self-gravity (that is, $\mu/R\ll1$) and
that it is much smaller than the geometric length-scale set by the
black-hole horizon radius. In addition, the weak (positive) energy
condition implies that the radius of the charged body is bounded
from below by its classical radius \cite{Cr1,Cr2,Notepos}
\begin{equation}\label{Eq7}
R\geq R_{\text{c}}\equiv {{q^2}\over{2\mu}}\  .
\end{equation}
This inequality ensures that the energy density inside the spherical
charged body is positive \cite{ElBek}.

The energy \cite{Noteen} of the charged body in the near-horizon
black-hole spacetime is given by \cite{Hodo,ElBek}
\begin{equation}\label{Eq8}
{\cal
E}(r)=\mu\sqrt{{{r^2_0-2Mr_0+a^2}\over{r^2_0+a^2}}}+{{Mq^2}\over{2(r^2_0+a^2)}}\
,
\end{equation}
where $r=r_0$ is the radial coordinate of the body's center of mass
in the black-hole spacetime. The first term on the r.h.s of
(\ref{Eq8}) represents the energy associated with the rest mass
$\mu$ of the body red-shifted by the black-hole gravitational field
\cite{Bek2,Car}. The second term on the r.h.s of (\ref{Eq8})
represents the self-energy of the charged body in the curved
black-hole spacetime \cite{Lon,ElBek,Hodo,Noteself}.

The proper height $l$ of the body's center of mass above the
black-hole horizon is related by the integral relation \cite{Bek2}
\begin{equation}\label{Eq9}
l(r_0)=\int_{r_+}^{r_0}\sqrt{{{r^2+a^2}\over{r^2-2Mr+a^2}}}dr\
\end{equation}
to the Boyer-Lindquist radial coordinate $r_0$. In the near-horizon
$l\ll r_+$ region one finds the relation
\begin{equation}\label{Eq10}
r_0(l)-r_+=(r_+-r_-){{l^2}\over{4\alpha}}[1+O(l^2/r^2_+)]\  ,
\end{equation}
where $\alpha\equiv r^2_++a^2$. Taking cognizance of Eqs.
(\ref{Eq8}) and (\ref{Eq10}), one finds
\begin{equation}\label{Eq11}
{\cal E}(l)={{(r_+-r_-)\mu
l+Mq^2}\over{2\alpha}}\cdot[1+O(l^2/r^2_+)]\
\end{equation}
for the energy of the body in the near-horizon $l\ll r_+$ region.

Suppose now that the charged object is slowly lowered towards the
black hole until its center of mass lies a proper height $l_0$ (with
$l_0\geq R$) above the black-hole horizon. The object is then
released to fall into the black hole. The assimilation of the
charged body by the black hole produces a final Kerr-Newman
black-hole configuration whose physical parameters (mass, charge,
and angular momentum) are given by
\begin{equation}\label{Eq12}
M\to M_{\text{new}}=M+{\cal E}(l_0)\ \ \ ; \ \ \ a\to
a_{\text{new}}=a[1-{\cal E}(l_0)/M+O({\cal E}^2/M^2)]\ \ \ ; \ \ \
Q=0\to Q_{\text{new}}=q\ .
\end{equation}
The change in the black-hole temperature caused by the assimilation
of the charged body can be quantified by the dimensionless physical
function
\begin{equation}\label{Eq13}
\Theta({\bar a})\equiv {{\Delta T_{\text{BH}}}\over{T_{\text{BH}}}}\
,
\end{equation}
where ${\bar a}\equiv a/M$ is the dimensionless angular momentum of
the black hole \cite{Noteci}.

Our goal is to bring the black hole as close as possible to its
extremal (zero-temperature) limit. Thus, we would like to minimize
the value of the dimensionless physical parameter $\Theta$. In
particular, we would like to examine whether $\Theta({\bar a})$, the
dimensionless change in the black-hole temperature, can be made {\it
negative} all the way down to the extremal ${\bar a}\to1$
(zero-temperature, $T_{\text{BH}}\to 0$) limit.

We shall henceforth consider black holes in the regime
\begin{equation}\label{Eq14}
{\bar a}\geq \sqrt{2\sqrt{3}-3}\  ,
\end{equation}
in which case a minimization of the energy delivered to the black
hole also corresponds to a minimization of the Bekenstein-Hawking
temperature of the final black-hole configuration \cite{Noteya}. The
fact that the energy ${\cal E}(l_0)$ of the charged particle in the
black-hole spacetime is an increasing function of the dropping
height $l_0$ [see Eq. (\ref{Eq11})] implies that, in order to
minimize the physical parameter $\Theta({\bar a})$ in the regime
(\ref{Eq14}), one should release the body to fall into the black
hole from a point whose proper height above the black-hole horizon
is as {\it small} as possible. We therefore face the important
question: How small can the dropping height $l_0$ be made?

As pointed out by Bekenstein \cite{Bek2}, the expression
(\ref{Eq11}) for the energy of our charged spherical object in the
black-hole spacetime is only valid in the restricted regime $l_0\geq
R$, where every part of the body is still {\it outside} the horizon.
This fact implies, in particular, that the adiabatic (slow) descent
of the charged spherical body towards the black hole must stop when
its center of mass lies a proper height $l_0\to R^+$ above the
horizon. At this point the bottom of the body is almost swallowed by
the black hole and the body [having a minimized (red-shifted) energy
${\cal E}(l_0\to R)$] should then be released to fall into the black
hole \cite{Bek2}. In addition, remembering that the weak (positive)
energy condition sets the lower bound (\ref{Eq7}) on the proper
radius of the charged spherical body, one finds the relation
\cite{Notef1}
\begin{equation}\label{Eq15}
l^{\text{min}}_0=R^{\text{min}}={{q^2}\over{2\mu}}\
\end{equation}
for the optimal \cite{Noteop} dropping point of the charged body.

Substituting (\ref{Eq15}) into (\ref{Eq11}), one finds the
remarkably simple (and universal \cite{Noteuni1}) expression
\begin{equation}\label{Eq16}
{\cal E}^{\text{min}}({\bar a})={{q^2}\over{4M}}
\end{equation}
for the minimal energy delivered to the black hole by the charged
body. Taking cognizance of Eqs. (\ref{Eq2}), (\ref{Eq12}),
(\ref{Eq13}), and (\ref{Eq16}), one finds the universal expression
\cite{Noteuni2,Notegn,Noteth}
\begin{equation}\label{Eq17}
\Theta^{\text{min}}({\bar a})=-{{q^2}\over{4M^2}}
\end{equation}
for the smallest possible (most negative) value of the dimensionless
physical parameter $\Theta({\bar a})$ which quantifies the change in
the black-hole temperature caused by the assimilation of the charged
body. Interestingly, one finds from (\ref{Eq17}) the characteristic
inequality
\begin{equation}\label{Eq18}
\Theta^{\text{min}}({\bar a})<0\  ,
\end{equation}
which is valid for {\it all} values ${\bar a}\in [0,1)$ of the
black-hole rotation parameter. The simple inequality (\ref{Eq18})
implies that, by absorbing charged particles, the black hole can
approach arbitrarily close to the extremal (zero-temperature)
$T_{\text{BH}}\to 0$ limit.

It is important to emphasize again that this conclusion is based on
the {\it assumption} \cite{Bek2} that the charged body can be
lowered adiabatically (slowly) until its bottom almost touches the
black-hole horizon \cite{Notelh}. In the next section we shall show,
however, that Thorne's famous hoop conjecture \cite{Thorne} implies
that, for near-extremal black holes, the charged body {\it cannot}
be lowered adiabatically all the way down to the horizon of the
black hole.

\section{The hoop conjecture and the lower bound on the black-hole temperature}

In the previous section we have seen that, by absorbing a charged
particle, a black hole can approach arbitrarily close to the
extremal (zero-temperature) $T_{BH}\to 0$ limit. As we have
emphasized above, this interesting conclusion rests on the
assumption that the charged body can be lowered slowly all the way
down to the horizon of the black hole \cite{Notelh}. In the present
section we shall show, however, that Thorne's famous hoop conjecture
\cite{Thorne} sets a lower bound on the minimal proper height
$l^{\text{min}}_0$ that the charged body can approach the black-hole
horizon without being absorbed, a bound which may be stronger than
the previously assumed bound (\ref{Eq15}).

The Thorne hoop conjecture \cite{Thorne} asserts that a physical
system of total mass (energy) $M$ forms a black hole if its
circumference radius $r_{\text{c}}$ is equal to (or smaller than)
the corresponding radius $r_{\text{Sch}}=2M$ of the Schwarzschild
black hole. It is worth emphasizing that the validity of this
version of the hoop conjecture is supported by several studies
\cite{Teuk}. However, it is also important to emphasize the fact
that there are known spacetime solutions of the Einstein field
equations
which provide explicit counterexamples to this version of the hoop
conjecture \cite{Leon,Hak}.

A weaker (and therefore a more robust) version of the hoop
conjecture for spacetimes with no angular momentum was suggested in
\cite{Hodm,Notenec}. Here we would like to generalize this weaker
version of the hoop conjecture to the generic case of spacetimes
which possess angular momentum and electric charge. In particular,
we conjecture that: A physical system of mass $M$, angular momentum
$J$, and electric charge $Q$ forms a black hole if its circumference
radius $r_{\text{c}}$ is equal to (or smaller than) the
corresponding Kerr-Newman black-hole radius
$r_{\text{KN}}=M+\sqrt{M^2-(J/M)^2-Q^2}$. That is, we conjecture
that
\begin{equation}\label{Eq19}
r_{\text{c}}\leq M+\sqrt{M^2-(J/M)^2-Q^2}\ \  \Longrightarrow \ \
\text{Black-hole horizon exists}\  .
\end{equation}

In the context of our gedanken experiment, this weaker version of
the hoop conjecture implies that a new (and larger) horizon is
formed if the charged body reaches the radial coordinate
$r_0=r_{\text{hoop}}$, where $r_{\text{hoop}}(\mu,q)$ is defined by
the Kerr-Newman functional relation [see Eq. (\ref{Eq3})]
\begin{equation}\label{Eq20}
r_{\text{hoop}}=M+{\cal E}(r_{\text{hoop}})+\sqrt{[M+{\cal
E}(r_{\text{hoop}})]^2-\{J/[M+{\cal
E}(r_{\text{hoop}})]\}^2-(Q+q)^2}\ .
\end{equation}
Substituting (\ref{Eq8}) into (\ref{Eq20}), and assuming
$r_{\text{hoop}}-r_+\ll r_+-r_-\ll r_+$ \cite{Notene}, one finds
\cite{Notenep}
\begin{equation}\label{Eq21}
r_{\text{hoop}}-r_+={{2\beta^2\mu^2}\over{r_+-r_-}}
\end{equation}
for the radius of the new horizon, where
\begin{equation}\label{Eq22}
\beta\equiv 1+\sqrt{1-{{q^2}\over{8\mu^2}}\cdot\tau}\ \ \ \
\text{with}\ \ \ \ \tau\equiv {{r_+-r_-}\over{r_+}}\  .
\end{equation}

Substituting the radial coordinate (\ref{Eq21}) into Eq.
(\ref{Eq10}), one finds
\begin{equation}\label{Eq23}
l(r_{\text{hoop}})={{4\beta\mu}\over{\tau}}\ .
\end{equation}
Taking cognizance of Eqs. (\ref{Eq15}) and (\ref{Eq23}) one realizes
that, in the regime
\begin{equation}\label{Eq24}
l(r_{\text{hoop}})>R^{\text{min}}={{q^2}\over{2\mu}}\  ,
\end{equation}
a new (and larger) horizon is formed \cite{Notenh} {\it before} the
spherical charged body \cite{Noterr} touches the horizon of the
original black hole. Thus, in the regime (\ref{Eq24}), one should
take \cite{Notef2}
\begin{equation}\label{Eq25}
l^{\text{min}}_0=l(r_{\text{hoop}})
\end{equation}
in Eq. (\ref{Eq11}) in order to minimize the energy delivered to the
black hole by the charged body \cite{Noteag}. This implies
\cite{Notenep}
\begin{equation}\label{Eq26}
{\cal E}^{\text{min}}({\bar a})={{4\beta{\mu}^2+q^2}\over{4r_+}}\
\end{equation}
for the smallest possible energy delivered by the charged particle
to the black hole in the regime (\ref{Eq24}). Taking cognizance of
Eqs. (\ref{Eq2}), (\ref{Eq12}), (\ref{Eq13}), and (\ref{Eq26}), one
finds the relation
\begin{equation}\label{Eq27}
\Theta^{\text{min}}({\bar a})={{{{8\beta {\mu}^2}\over{\tau}}-
q^2}\over{2r^2_+\sqrt{1-{\bar a}^2}}}\
\end{equation}
in the regime (\ref{Eq24}).

Interestingly, one finds from (\ref{Eq27}) that the
black-hole-charged-body system is characterized by the inequality
\begin{equation}\label{Eq28}
\Theta^{\text{min}}({\bar a})>0\
\end{equation}
in the regime (\ref{Eq24}). Note, in particular, that the inequality
(\ref{Eq24}) is satisfied by near-extremal black holes whose
dimensionless temperature $\tau$ is characterized by the relation
[see Eqs. (\ref{Eq22}) and (\ref{Eq23})]
\begin{equation}\label{Eq29}
\tau<{{8\mu^2}\over{q^2}}\ .
\end{equation}
Taking cognizance of Eqs. (\ref{Eq28}) and (\ref{Eq29}) one realizes
that, in our gedanken experiment, the Bekenstein-Hawking temperature
of the black holes {\it cannot} be lowered below the critical value
\cite{Notetu}
\begin{equation}\label{Eq30}
T^{\text{c}}_{\text{BH}}\times r_+={{\hbar}\over{\pi}}\cdot
{{\mu^2}\over{q^2}} \ ,
\end{equation}
where $\mu$ and $q$ are the proper mass and electric charge of the
absorbed particle, respectively.

\section{The quantum buoyancy effect and the lower bound on the black-hole temperature}

Thus far, we have analyzed the gedanken experiment at the {\it
classical} level. It is important to emphasize, however, that the
well known {\it quantum} buoyancy effect \cite{UWa} in the
black-hole spacetime should also be taken into account in the
present gedanken experiment. This quantum buoyancy effect stems from
the fact that the {\it slowly} lowered object interacts with the
quantum thermal atmosphere of the black-hole spacetime
\cite{UWa,Bekw}.

In particular, as shown by Bekenstein \cite{Bekw}, the quantum
buoyancy effect shifts the optimal dropping point \cite{Noteop} of
the object from $l^{\text{min}}_0=R$ [see Eq. (\ref{Eq15})] to a
slightly higher point whose proper radial distance from the
black-hole horizon is given by \cite{Bekw}
\begin{equation}\label{Eq31}
l^{\text{min}}_0=(1+\epsilon)\cdot R\  ,
\end{equation}
where the dimensionless factor $\epsilon$ is given by
\cite{Bekw,Notesz}
\begin{equation}\label{Eq32}
\epsilon\equiv \sqrt{{{N}\over{720\pi}}\cdot {{\hbar}\over{\mu R}}}
\end{equation}
and $N$ is the effective number of quantum radiation species
\cite{Bekw}. The quantum shift ({\it increase}) $\epsilon R$ [see
Eq. (\ref{Eq31})] in the radial proper distance of the optimal
dropping point results in a quantum {\it increase}
$\epsilon\cdot(r_+-r_-)\mu R/\alpha$ \cite{Bekw} in the energy
delivered to the black hole. Taking into account this quantum
buoyancy increase in the energy delivered to the black hole, one
finds that the classical expression (\ref{Eq17}) for the
dimensionless function $\Theta({\bar a})$ acquires a {\it positive}
quantum correction term. In particular, for near-extremal black
holes the quantum-mechanically corrected expression for
$\Theta({\bar a})$ is given by \cite{Noterr,Notenep}
\begin{equation}\label{Eq33}
\Theta^{\text{min}}({\bar
a}\to1)=-{{q^2}\over{4M^2}}\cdot\Big(1-\epsilon\cdot{{8r_+}\over{r_+-r_-}}\Big)\
.
\end{equation}

Interestingly, one finds from (\ref{Eq33}) that the
black-hole-charged-body system is characterized by the inequality
\begin{equation}\label{Eq34}
\Theta^{\text{min}}({\bar a})>0\
\end{equation}
in the regime
\begin{equation}\label{Eq35}
\tau<8\epsilon\ .
\end{equation}
The relations (\ref{Eq34}) and (\ref{Eq35}) imply that, due to the
{\it quantum} buoyancy effect \cite{UWa,Bekw}, the
Bekenstein-Hawking temperature of the black holes {\it cannot} be
lowered below the critical value \cite{Notetu}
\begin{equation}\label{Eq36}
T^{\text{c}}_{\text{BH}}\times r_+=\epsilon\cdot{{\hbar}\over{\pi}}\
.
\end{equation}
Interestingly, the quantum lower bound (\ref{Eq36}) becomes stronger
than the classical lower bound (\ref{Eq30}) in the regime
$\epsilon>\mu^2/q^2$, which corresponds to charged objects
\cite{Noterr} in the regime $q>(360\pi/N\hbar)^{1/2}\mu^2$.

\section{Summary and discussion}

We have analyzed a gedanken experiment in which a spherical charged
particle is lowered into a Kerr black hole. It was shown that {\it
if} the charged particle can be lowered slowly all the way down to
the horizon of the black hole, then the Bekenstein-Hawking
temperature of the final black-hole configuration can approach
arbitrarily close to the extremal (zero-temperature)
$T_{\text{BH}}\to0$ limit.

However, we have shown that Thorne's famous hoop conjecture
\cite{Thorne} [and also its weaker (and more robust) generalization
(\ref{Eq19})] implies that, for near-extremal black holes in the
regime (\ref{Eq29}), a new (and larger) horizon is already formed
{\it before} the charged particle touches the horizon of the
original black hole. The hoop conjecture therefore implies that, in
our gedanken experiment, the temperature of the final \cite{Notefn}
black-hole configuration {\it cannot} approach arbitrarily close to
zero \cite{Noteoh}. In particular, we have proved that the
Bekenstein-Hawking temperature of the black holes is an {\it
irreducible} quantity in the near-extremal regime
$T_{\text{BH}}<T^{\text{c}}_{\text{BH}}$ determined by the critical
temperature (\ref{Eq30}).

It is worth emphasizing that we have provided in this paper only one
specific example, {\it not} a general proof, to the fact that the
black-hole temperature cannot approach arbitrarily close to zero.
Nevertheless, this intriguing conclusion of our gedanken experiment
\cite{Notesm} makes it tempting to conjecture that the
Bekenstein-Hawking temperature of black holes is bounded from below
by the simple universal relation [see Eq. (\ref{Eq30})]
\cite{Notesi,Noteul,Noteqb}
\begin{equation}\label{Eq37}
T_{\text{BH}}\times r_+\gg \Big({{\hbar}\over{r_+}}\Big)^2\ .
\end{equation}
We believe that it would be highly important to test the general
validity of the conjectured lower bound (\ref{Eq37}) on the
Bekenstein-Hawking temperature of the black holes.

\bigskip
\noindent
{\bf ACKNOWLEDGMENTS}
\bigskip

This research is supported by the Carmel Science Foundation. I would
like to thank Yael Oren, Arbel M. Ongo, Ayelet B. Lata, and Alona B.
Tea for helpful discussions.


\begin{thebibliography}{99}
\bibitem{Noteunit} We shall use gravitational units in which $G=c=k_{\text{B}}=1$.
\bibitem{Notema} Here $M$, $J\equiv Ma$, and $Q$ are respectively the mass, angular
momentum, and electric charge of the Kerr-Newman black hole.
\bibitem{Notethr} It is worth emphasizing that there are several ways to define the effective size
of a black-hole system. In particular, the black-hole spacetime is
characterized by an infinite throat in the extremal limit. This
infinite throat may suggest that the effective size of the
black-hole system diverges in the extremal (zero-temperature) limit.
\bibitem{Notegen} We shall assume $q>0$ and $a>0$ without loss of
generality.
\bibitem{Notepos} Note that the limiting case $R=R_{\text{c}}=q^2/2\mu$ corresponds to
a charged spherical object whose rest mass $\mu$ is attributed
exclusively to the electrostatic self-energy it produces
\cite{Cr1,Cr2}. The factor of $1/2$ in (\ref{Eq7}) comes from the
assumption that the electric charge is uniformly spread on a thin
spherical shell.
\bibitem{Noteen} We refer here to the energy as measured by
asymptotic observers.
\bibitem{Noteself} The physical origin of this self-interaction $O(q^2/M)$
term is attributed to the distortion of the body's long-range
Coulomb field by the curved black-hole spacetime
\cite{Lon,ElBek,Hodo}.
\bibitem{Noteci} We recall that the parameters $\{M,a\}$ are the mass and angular momentum
per unit mass of the original Kerr black hole.
\bibitem{Noteya} That is, Kerr black holes in the regime (\ref{Eq14})
are characterized by the relation $({\partial
T_{\text{BH}}}/{\partial M})_{J,Q}>0$ [see Eqs. (\ref{Eq2}) and
(\ref{Eq3})].
\bibitem{Notef1} It is important to emphasize that our assumption $l^{\text{min}}_0\ll r_+$ [see Eq. (\ref{Eq10})]
corresponds to charged particles in the regime $q^2\ll\mu r_+$.
\bibitem{Noteop} That is, the dropping point for which the energy delivered to the black hole,
and thus also the physical parameter $\Theta({\bar a})$, are {\it
minimized}.
\bibitem{Noteuni1} The expression (\ref{Eq16}) for the minimal energy delivered to the
black hole by the charged body is universal in the sense that it is
{\it independent} of the black-hole rotation parameter ${\bar a}$.
\bibitem{Notegn} It is worth noting that, for generic values of the dropping
height $l_0$ and in the regime $q^2\ll r_+(r_+-r_-)$, one finds from
Eqs. (\ref{Eq2}), (\ref{Eq11}), (\ref{Eq12}), and (\ref{Eq13}), the
relation $\Theta(l_0;{\bar a})={{\big(1+{\bar a}^2-2\sqrt{1-{\bar
a}^2}\big)\cdot 2\mu l_0-\big(2-\sqrt{1-{\bar a}^2}\big)\cdot
q^2}\over{2\alpha\sqrt{1-{\bar a}^2}}}$ for the dimensionless
physical parameter $\Theta$ which quantifies the change in the
black-hole temperature caused by the absorbed particle. This
expression implies that, in the regime (\ref{Eq14}), the
dimensionless physical parameter $\Theta(l_0;{\bar a})$ is an
increasing function of the dropping height $l_0$ (see
\cite{Noteya}).
\bibitem{Notenec} Interestingly, it was shown in \cite{Hodm} that the (weaker version \cite{Hodm} of the) hoop
conjecture must be invoked in order to guarantee the validity of the
generalized second law of thermodynamics \cite{Bek2} in a gedanken
experiment in which an entropy bearing object is lowered slowly into
a near-extremal black hole.
\bibitem{Notene} We shall henceforth assume that the original Kerr black
hole is a near-extremal one.
\bibitem{Notenep} Here we have used the approximated relations
${\bar a}\simeq 1$ and $\alpha\simeq 2r^2_+$ for near-extremal black
holes with $a\simeq M\simeq r_+$.
\bibitem{Notenh} As discussed above, this new (and larger)
horizon is expected to be formed according to the original hoop
conjecture \cite{Thorne} and its generalized version (\ref{Eq19}).
\bibitem{Noterr} We recall that the proper radius of the charged spherical body
is given by $R=R^{\text{min}}=q^2/2\mu$ [see Eq. (\ref{Eq7})].
\bibitem{Notef2} It is important to emphasize that our assumption $l^{\text{min}}_0\ll r_+$
corresponds to particles in the regime $\mu\ll r_+-r_-$ [see Eq.
(\ref{Eq23})].
\bibitem{Noteag} It is worth emphasizing again that, in the regime (\ref{Eq14}), a minimization of
the energy (\ref{Eq11}) which is delivered to the black hole also
corresponds to a minimization of the dimensionless physical
parameter $\Theta$ which quantifies the change in the black-hole
temperature (see \cite{Noteya} and \cite{Notegn}).
\bibitem{Notetu} Here we have used the approximated relation $T_{\text{BH}}\times r_+\simeq \tau\hbar/8\pi$
for the near-extremal Kerr-Newman black holes with ${\bar
a}\simeq1$.
\bibitem{Notesz} As shown by Bekenstein \cite{Bekw}, the relations (\ref{Eq31}) and (\ref{Eq32}) are
valid for macroscopic and mesoscopic bodies in the regime $R\gg
\hbar/\mu$. This strong inequality corresponds to the regime
$\epsilon\ll1$.
\bibitem{Notefn} That is, after the assimilation of the charged
particle by the black hole.
\bibitem{Noteoh} It is worth noting that, had we used the
original hoop conjecture \cite{Thorne} instead of its weaker version
(\ref{Eq19}), we would have found that the Bekenstein-Hawking
temperature of the final Kerr-Newman black-hole configuration (after
the assimilation of the charged particle) is higher than the
temperature of the original Kerr black hole in the entire regime
${\bar a}\in [0,1]$.
\bibitem{Notesi} Here we have used the relation $\mu^2/q^2=\mu/2R$
[see Eq. (\ref{Eq7})] for our charged massive particle. In addition,
we have used the inequalities $\hbar/\mu\leq R\ll r_+$ [see Eq.
(\ref{Eq6})] which characterize the physical parameters of the
captured particle. (As emphasized by Bekenstein \cite{Bek2}, the
inequality $R\geq\hbar/\mu$ reflects the fact that the proper radius
of the particle is bounded from below by its Compton length
\cite{Bek2}).
\end{thebibliography}
\end{document}